\documentclass[twoside]{dis07}
\usepackage[latin1]{inputenc}
\usepackage[dvips]{graphicx,epsfig,color}
\usepackage{wrapfig,rotating}
\usepackage{cite}
\usepackage{amssymb,amsmath,array}

\newcommand\GeV{\,\mbox{${\rm GeV}$}\,}
\newcommand\MeV{\,\mbox{${\rm MeV}$}\,}

\begin{document}
\title{{\normalsize\sl DESY 07--114     \hfill {\tt arXiv:0708.1474}
\\
\vspace*{-2mm}
SFB/CPP-07-48 \hfill {   } }\\
The Status of the Polarized Parton Densities}

\author{Johannes Bl\"umlein 
%
\thanks{This paper was supported in part by SFB-TR-9: Computergest\"utze 
Theoretische Teilchenphysik.}
%
\vspace{.3cm}\\
%
Deutsches Elektronen-Synchrotron, DESY,
Platanenallee 6, D-15738 Zeuthen, Germany
}

\maketitle

\begin{abstract}
A survey is given on the present knowledge of the polarized parton 
distribution functions. We give an outlook on further developments
desired both on the theoretical as well on the experimental side to
complete the understanding of the spin--structure of nucleons in the 
future.
\end{abstract}

\section{Introduction}

\vspace{1mm} \noindent
Deeply inelastic scattering provides a clean way to extract the parton 
densities of nucleons. After the initial observation that the nucleon spin 
is not formed by the quarks dominantly \cite{SPUZ}, detailed measurements 
of the polarized structure functions followed during the last 
20~years. The central question concerns now the parton distribution functions 
and their scale evolution rather than just their first moment. Since the 
nucleon spin receives also contributions from the angular momentum of 
the quarks and gluons, these degrees of freedom have also to be studied.
This requires the analysis of non--forward scattering cross sections. In 
inclusive deep--inelastic scattering the sensitivity to resolve the 
different sea--quark contributions is rather limited. One way to extract
this information consists in measuring semi--inclusive processes 
\cite{SEEX}. A central question concerns the polarized gluon distribution,
which can be determined from the scaling violations of $g_1(x,Q^2)$, 
deep--inelastic heavy flavor production, and hard processes measured at 
hadron colliders. The inclusive and semi--inclusive hard processes in 
polarized scattering provide an important laboratory to test QCD. More 
than exploring the level of twist--2, which has been  
investigated in some detail already, one may probe the twist--3 contributions in 
various transverse spin processes.   
In the following we survey the theoretical status of inclusive polarized deeply 
inelastic scattering and the status of polarized parton densities, 
cf. also~\cite{Blumlein:2005wf}. We close with an outlook on investigations 
required in the future.
\section{Theoretical Aspects}

\vspace{1mm} \noindent
In the deep--inelastic domain the polarized nucleon structure functions 
receive contributions of leading and higher twist, depending on the
region in $Q^2$ and $W^2$ probed. The leading twist contributions are those of 
twist $\tau = 2$ for $g_1(x,Q^2)$~\footnote{
Note that $g_1(x,Q^2)$ contains twist $\tau = 3$ contributions due 
to target mass corrections,~\cite{BT}.} 
and $\tau = 2, 3$ for $g_2(x,Q^2)$. At the level of twist--2 one may extract the 
polarized parton densities from the data on the  structure 
function $g_1(x,Q^2)$ performing an analysis in the framework of perturbative 
QCD. During the past decades higher orders have been approached steadily. The running of
$\alpha_s(\mu^2)$ is known to $O(\alpha_s^4)$ \cite{ALP}, both the polarized anomalous dimensions 
\cite{ANDI}~\footnote{Due to the Ward identity $P^{qq}_{\rm NS} =  \Delta P^{qq}_{\rm NS}$ this 
splitting function is also known to $O(\alpha_s^3)$,~\cite{ANDI3}.} and massless Wilson 
coefficients \cite{WIL1} were calculated to $O(\alpha_s^2)$ and the first non-singlet moment, the polarized 
Bjorken sum-rule, to 
$O(\alpha_s^3)$ \cite{BSR}.
The heavy flavor Wilson coefficients, in the whole kinematic region, are
only known to $O(\alpha_s)$~\cite{HEAV1}. For $Q^2 \gg m^2$, i.e. in the region $Q^2/m^2 \gtrapprox 10$, 
the Wilson coefficients were calculated in $O(\alpha_s^2)$,~\cite{HEAV2}. 
An interesting property is exhibited by the gluonic heavy flavor Wilson coefficient, the first moment of which 
vanishes in leading and next-to-leading order \cite{HEAV1,HEAV2}. Given a positive polarized gluon density, 
this implies a negative correction to $g_1(x,Q^2)$ in the region $x \lessapprox 10^{-2}$ and a positive 
contribution 
above. Conversely, the Wandzura--Wilczek relation implies a positive correction for the twist--2 
part of $g_2(x,Q^2)$ for $x \lessapprox 
2 \cdot 10^{-2}$, but a negative correction for larger $x$--values, cf.~\cite{BRN2}. The anomalous 
dimensions for the evolution
of the transversity distribution are known to $O(\alpha_s^2)$ in general \cite{TRAN1} and for a series
of moments to $O(\alpha_s^3)$ \cite{TRAN2}. At present only next-to-leading order 
QCD analyzes can be performed to extract the polarized parton distributions.

At the level of the twist--3 contributions to the polarized structure functions several sum--rules
and integral relations were derived, cf.~\cite{BT,BK1}. The twist--2 contributions to the structure 
function $g_2(x,Q^2)$ is given by the Wandzura--Wilczek relation \cite{WW} for the quarkonic, gluonic, 
heavy flavor contributions, target mass corrections, and even diffractive scattering, 
cf.~\cite{WW1,BRN2,BT}.
The twist--3 contributions to the structure function $g_2(x,Q^2)$ were calculated to one--loop order.
The $O(\alpha_s)$ non-singlet and singlet anomalous dimension matrices, respectively 
their corresponding expressions in momentum fraction space,  were derived in 
Refs.~\cite{THR_ANOM} using different techniques. The $O(\alpha_s)$ Wilson 
coefficients were calculated in \cite{THR_WIL1}. Although the precision on 
$g_2(x,Q^2)$ improved during the last years \cite{Zheng:2004ce} and some difference between the
data and the Wandzura--Wilczek approximation is seen, still more precise data are required 
to test the QCD--predictions. First non-singlet moments of the twist--3 operator expectation values 
were determined in lattice simulations \cite{LATT1,NEG}. 

Also in case of the polarized structure functions small-$x$ resummations 
are discussed, which can be described on the basis of infrared evolution 
equations \cite{KL} and emerge both for the non-singlet \cite{BV1} and 
singlet structure functions \cite{BER,BV2}. These resummations apply to 
the leading pole $O((\alpha_s/N^3)^k)$-terms only, with $N$ the Mellin-variable. 
Performing the resummation one obtains a branch--cut instead of the perturbative pole-terms,
which yields a milder singularity.
The resummation is 
consistently 
accounted for by the Callan-Symanzik equations for the evolution of parton 
densities. As detailed numerical studies, which were performed in Refs.~\cite{BV1,BV2}, show~\footnote{
For the unpolarized case see \cite{BV3}.}, one has to take into account not only 
the leading pole terms, but also the {\sf resummed} sub-leading terms, see also \cite{BN1}, which 
are not yet calculated completely. 
They are known, however, for the first two 
orders in $\alpha_s$ (and partly to $O(\alpha_s^3)$)
for {\sf all} sub-leading terms, which suggest the general form.
 The comparison of the leading and sub-leading terms
shows, that at least three sub-leading terms are required to match the exact results.
Ignorance of these contributions,
as unfortunately still partly present in the contemporary literature, results into  misleading
quantitative analyzes. Sometimes also "DGLAP" evolution is opposed to 
"infrared evolution equations", etc. Here again a clarifying word is in order. 
In practice both concepts address twist--2 parton distributions. 
Their scale evolution results from the 
factorization of the collinear singularities and is
ruled by the anomalous dimensions $P_{ij}(N,a_s) = \sum_{k=1}^{\infty} a_s^k 
P_{ij}^{(k-1)}(N)$~.
The corresponding Callan-Symanzik equations have to be solved for high enough 
orders in the coupling constant
in the range of Bjorken-$x$, demanded by the experimental data. 
These equations cover the small-$x$ and the less singular terms which are equally important in
quantitative analyzes.
\section{Parton Distributions}

\vspace{1mm} \noindent
The polarized parton distribution functions may be determined by a QCD--analysis 
of the structure function $g_1(x,Q^2)$. The data analysis requires a detailed 
description of the denominator of the polarization asymmetry, which has to include
empiric parameterizations both for $F_2(x,Q^2)$ and $F_L(x,Q^2)$ including potential 
higher twist contributions, since the region of $Q^2$ and $W^2$, which is analyzed, 
covers rather low values, unlike the case in unpolarized analyzes~\cite{UNPOL}. 
Usually one would like to limit the data analysis to the region $Q^2 \gtrapprox 4 \GeV^2$, 
which will be possible in future measurements at a facility like EIC~\cite{EIC}.
The present data sets only allow a cut $Q^2 \gtrapprox 1 \GeV^2$. In the analysis 
the correlation of the different parameters of the parton distribution functions at $Q_0^2$
and the QCD--scale $\Lambda_{\rm QCD}$ are rather essential. Measuring the gluon distribution function $\Delta 
G(x,Q^2)$,
and to some extent also the sea--quark distributions, the slope effects of
$\partial g_1(x,Q^2)/\partial \ln(Q^2)$ are important. In case of $\Delta G$ there one observes a very
strong correlation with $\alpha_s(Q^2)$ due to the evolution equations. Special assumptions
on $\Lambda_{\rm QCD}$, as fixing this value to other measurements, may introduce severe biases.
In the inclusive analysis not all the parameters chosen to model  the parton distributions can be measured.
For the data sets currently available this applies in particular to those parameters which describe the
range of medium values of $x$. Their error may be- 

\vspace{1mm}
\hspace{1.4cm}
\begin{minipage}[h]{0.27\linewidth}
\hspace*{-3mm}
\centering\epsfig{figure=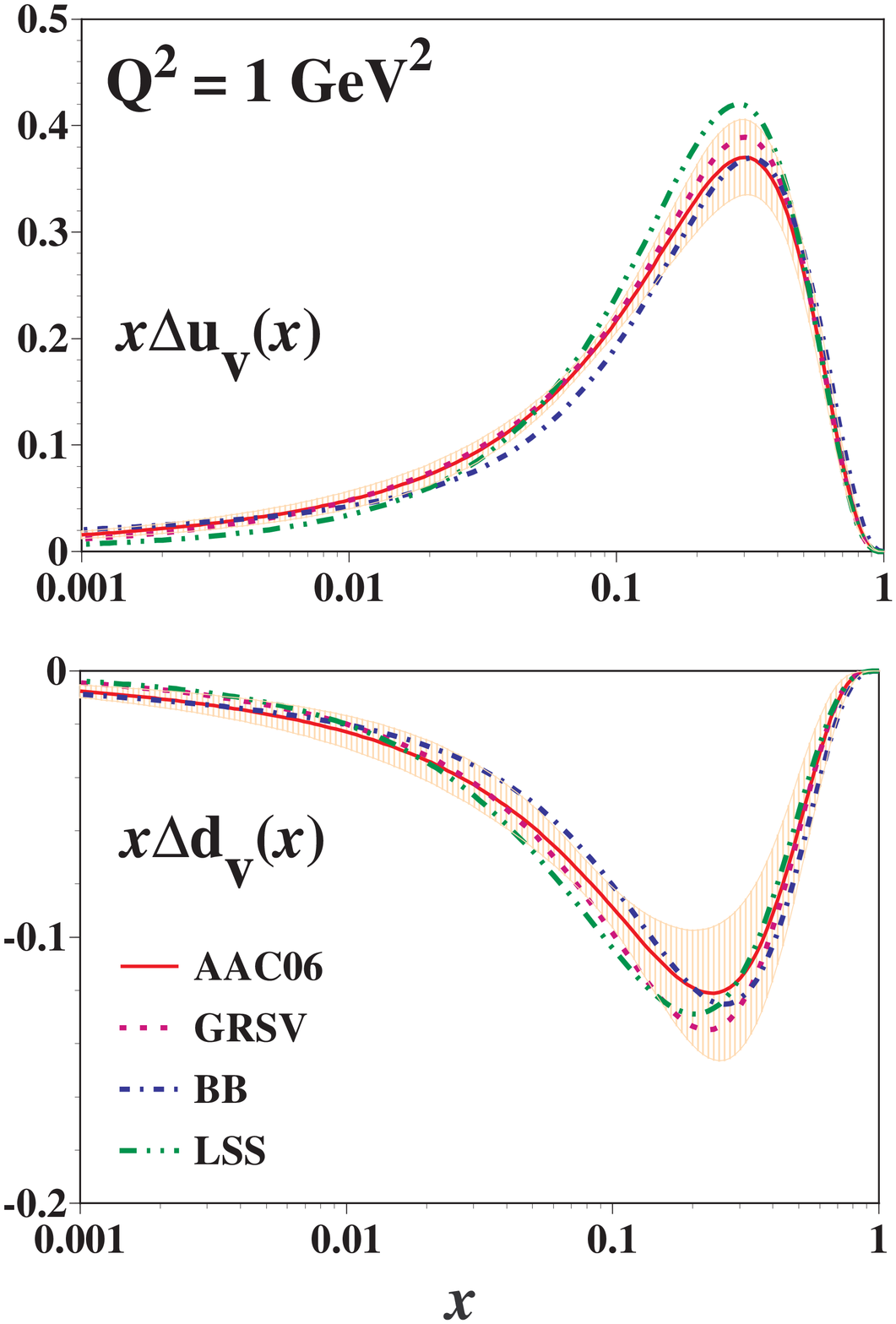,width=1.2\linewidth}\\
\end{minipage} \hspace{7mm} 
\begin{minipage}[h]{0.27\linewidth}
\centering\epsfig{figure=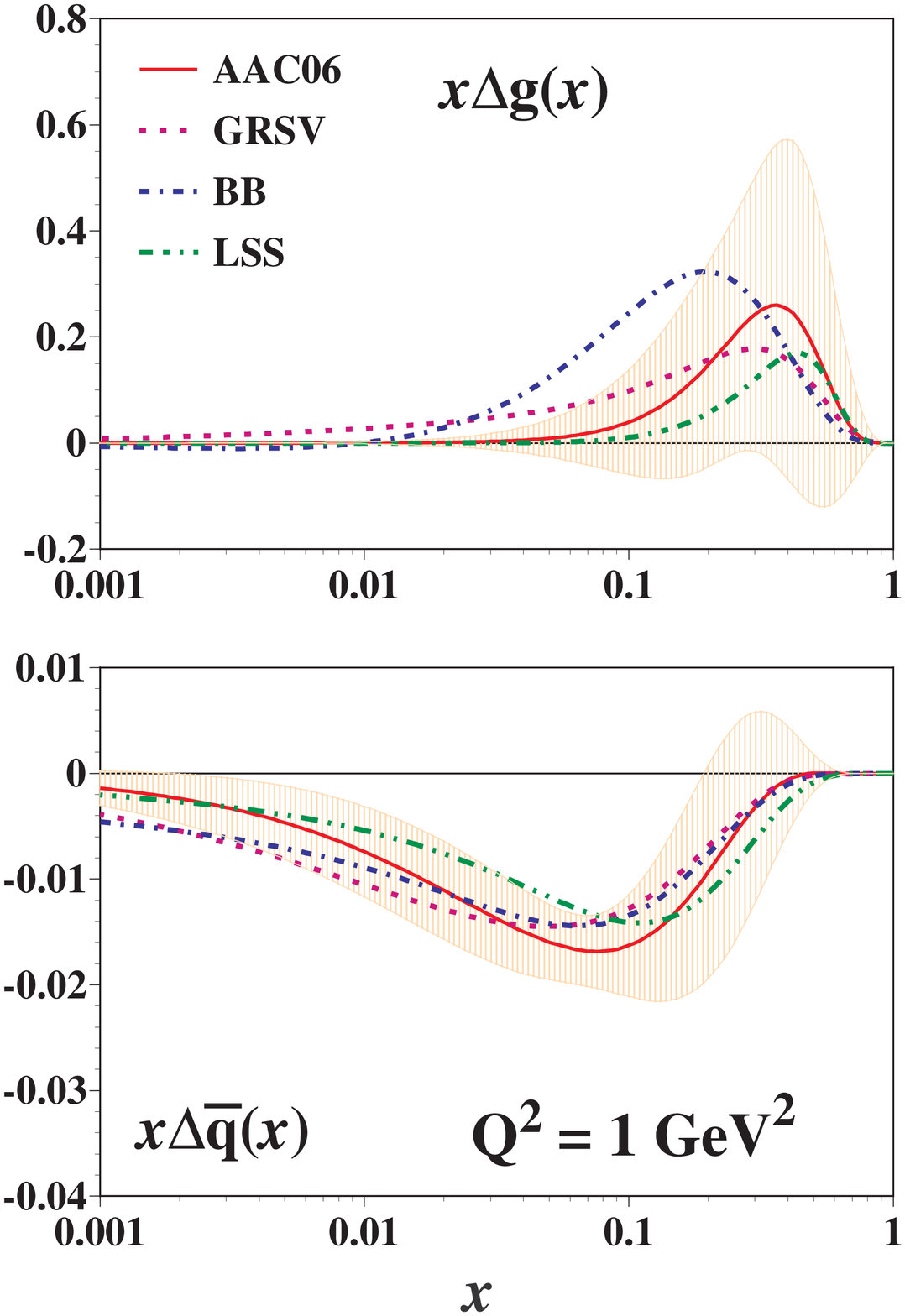,width=1.2\linewidth}
\end{minipage}

\vspace{1mm} \noindent
{\sf Figure~1: The polarized parton distributions from different analyzes at the scale $Q^2 = 1 
\GeV^2$,~\cite{AAC}. AAC06~:~\cite{AAC}; GRSV~:~\cite{GRSV}; BB~:~\cite{BB2}; LSS~:~\cite{LSS}.}

\vspace{1mm}\noindent
come rather large compared to their value. These
parameters can be fixed after a first minimization and form a certain model. Their value has to be re--fitted
after the global minimum was found, but will usually not change significantly, cf.~\cite{BB2}. 
The relative normalization of the different data sets is fitted within the allowed margins quoted by the 
experiments, to account for 
global systematic effects. In Figure~1 recent parton distribution functions \cite{BB2,AAC,GRSV,LSS} are 
compared. Further parameterizations were given in \cite{FNS,AK,NEUROL,COMP1}. Within the present errors
there is good agreement between all analyzes. The valence quark distribution functions are determined best,
with a positive polarized up-quark and and a negative down-quark distribution. The sea--quark distribution is 
found to be mainly negative, but with a larger error. To resolve the different flavors of the sea-quark 
contributions semi-inclusive data were analyzed \cite{SEEX}, yet with rather large errors. Under certain 
assumptions the error on the strange-quark density becomes rather low \cite{FNS}. Alternatively to the 
standard QCD fits neural-network techniques were used to determine the polarized parton 
densities in \cite{NEUROL}. In this analysis a larger error than found using the conventional methods is 
obtained 
in the small $x$ region, were data are sparse. If compared to earlier analyzes \cite{BB2} the polarized gluon 
distribution comes out at lower values including more recent data \cite{AAC,LSS,BB6}. The polarized gluon 
distribution function, although being obtained with positive central values in unconstrained fits, is still 
compatible with zero within the errors. In some analyzes \cite{COMP1,LSS,KUREK1} one demands, as second 
option, also a negative 
normalization of the gluon distribution in a constrained fit. The corresponding distribution is slightly 
negative and allowed by the data under the constraint used. The ratio $\Delta G/G$ was also measured in
open charm photo--production \cite{DELTA_G}. Here the experimental errors are still large and the result
is compatible with zero. Using the fit results of the polarized parton distributions one may form moments, cf. 
\cite{BB2}, to be compared with lattice simulations, in particular for the valence-quark distributions. Here 
the crucial point is to control the systematic 
effects and to approach realistic values of the pion mass. Currently values in the range $m_{\pi} \sim 270 
\MeV$ can be reached in dynamical quark simulations. In this way new non-perturbative quantitative test of 
QCD will be possible soon \cite{LATT1,NEG,LATT2}.   

\section{Future Avenues}

\vspace{1mm} \noindent
The current picture of the polarized nucleon is still in a move and more
efforts in theory and experiment are needed to complete it. In the 
forthcoming years the data-analysis from HERMES and COMPASS will lead to 
still better parton distribution functions. It would be important to 
measure the structure function $F_2(x,Q^2)$ at HERMES,
which would yield an improved systematic understanding of the data. Yet the
experimental precision of the structure function $g_1(x,Q^2)$ is limited 
and high--luminosity measurements at a future facility as EIC~\cite{EIC}  
is highly desirable to provide detailed QCD--tests for $\Lambda_{\rm 
QCD}$, the parton distributions and their moments to be compared with 
lattice simulations. The experiments at RHIC will improve our knowledge on 
the polarized sea--quark and gluon distributions. Important information on 
the large-$x$ behaviour of all distribution functions, can be gained in 
the experiments at JLAB running at an increased beam energy. The HERA 
experiments, COMPASS and the JLAB experiments will finalize their 
measurements on deeply-virtual Compton scattering (DVCS) \cite{DVCS} 
during the 
forthcoming 
years and we may hope to get constraints on the quark angular 
momentum \cite{LQ} using Ji's sum-rule \cite{JI}.

As shown in \cite{BB1} the theory error of the polarized gluon 
distribution at NLO is still large. The calculation of the 3--loop 
anomalous dimensions are therefore required to diminish this  
uncertainty. Very high statistics measurements have to be performed in the 
long--term future for DVSC to extract constraints on the angular momentum 
of the gluon from the scaling violations of the non--forward scattering 
cross sections. As is well--known, the scaling violations of the 
transversity distribution function $h_1(x,Q^2)$ are larger than those of 
the non--singlet part of $g_1(x,Q^2)$. Detailed high statistics 
measurements of this quantity are desirable to establish this 
prediction of QCD experimentally. Very little is known about the higher twist 
contributions to polarized deeply--inelastic scattering. Here we may hope 
for results from the experiments at JLAB. For the general kinematic region again 
high--luminosity experiments as planned for EIC would provide an excellent 
opportunity. Dedicated studies of the twist--3 contributions to 
several processes should be carried out and measurements shall be 
performed to isolate the twist--4 contributions. An interesting open issue is 
formed by twist--3 effects \cite{BK1,BT} in deep--inelastic scattering in 
electro--weak interactions, which can be studied  at future neutrino factories \cite{NUFACT}.
The present status of our knowledge on polarized parton densities is not 
yet sufficient and calling for refined measurements in various 
places which are crucial for the final understanding of the spin--structure of 
the nucleons. This will require extensive experimentation at a 
high--luminosity facility such as the future Electron--Ion--Collider.

%
%
%
%
\begin{footnotesize} 

\end{footnotesize}

\begin{thebibliography}{99} 
%
\bibitem{SPUZ}
  M.~J.~Alguard {\it et al.} (SLAC),
  Phys.\ Rev.\ Lett.\  {\bf 37} (1976) 1261;
{\bf 41} (1978) 70.
%
\bibitem{SEEX}
  A.~Airapetian {\it et al.}  [HERMES Collaboration],
  Phys.\ Rev.\  D {\bf 71} (2005) 012003.
%
\bibitem{Blumlein:2005wf}
  J.~Bl\"umlein,
  arXiv:hep-ph/0510212.
%
\bibitem{BT}
  J.~Bl\"umlein and A.~Tkabladze,
  Nucl.\ Phys.\  B {\bf 553} (1999) 427.
%
\bibitem{ALP}
  T.~van Ritbergen, J.~A.~M.~Vermaseren and S.~A.~Larin,
  Phys.\ Lett.\  B {\bf 400} (1997) 379;\\
  M.~Czakon,
  Nucl.\ Phys.\  B {\bf 710} (2005) 485.
%
\bibitem{ANDI}
  R.~Mertig and W.~L.~van Neerven,
  Z.\ Phys.\  C {\bf 70} (1996) 637;\\
  W.~Vogelsang,
  Nucl.\ Phys.\  B {\bf 475} (1996) 47;
  Phys.\ Rev.\  D {\bf 54} (1996) 2023.
%
\bibitem{ANDI3}
  S.~Moch, J.~A.~M.~Vermaseren and A.~Vogt,
  Nucl.\ Phys.\  B {\bf 688} (2004) 101.
%
\bibitem{WIL1}
  E.~B.~Zijlstra and W.~L.~van Neerven,
  Nucl.\ Phys.\  B {\bf 417} (1994) 61
  [Erratum-ibid.\  B {\bf 426} (1994) 245; {\bf 773} (2007) 105].
%
\bibitem{BSR}
  S.~A.~Larin and J.~A.~M.~Vermaseren,
  Phys.\ Lett.\  B {\bf 259} (1991) 345.
%
\bibitem{HEAV1}
  A.~D.~Watson,
  Z.\ Phys.\  C {\bf 12} (1982) 123.
%
\bibitem{HEAV2}
  M.~Buza, Y.~Matiounine, J.~Smith and W.~L.~van Neerven,
  Nucl.\ Phys.\  B {\bf 485} (1997) 420;\\
J. Bl\"umlein and S. Klein,  DESY 07--027;\\
  I.~Bierenbaum, J.~Bl\"umlein and S.~Klein,
  arXiv:0706.2738 [hep-ph]. 
%
\bibitem{BRN2}
  J.~Bl\"umlein, V.~Ravindran and W.~L.~van Neerven,
  Phys.\ Rev.\  D {\bf 68} (2003) 114004.
%
\bibitem{TRAN1}
  S.~Kumano and M.~Miyama,
  Phys.\ Rev.\  D {\bf 56} (1997) 2504;\\
  A.~Hayashigaki, Y.~Kanazawa and Y.~Koike,
  Phys.\ Rev.\  D {\bf 56} (1997) 7350;\\
  W.~Vogelsang,
  Phys.\ Rev.\  D {\bf 57} (1998) 1886.
%
\bibitem{TRAN2}
  J.~A.~Gracey,
  Phys.\ Lett.\  B {\bf 643} (2006) 374;
  JHEP {\bf 0610} (2006) 040.
%
\bibitem{BK1}
  J.~Bl\"umlein and N.~Kochelev,
  Nucl.\ Phys.\  B {\bf 498} (1997) 285;
%
\bibitem{WW}
  S.~Wandzura and F.~Wilczek,
  Phys.\ Lett.\  B {\bf 72} (1977) 195.
%
\bibitem{WW1}
  J.~D.~Jackson, G.~G.~Ross and R.~G.~Roberts,
  Phys.\ Lett.\  B {\bf 226} (1989) 159;\\
  J.~Bl\"umlein and N.~Kochelev,
  Phys.\ Lett.\  B {\bf 381} (1996) 296;\\
  A.~Piccione and G.~Ridolfi,
  Nucl.\ Phys.\  B {\bf 513} (1998) 301;\\
  J.~Bl\"umlein and D.~Robaschik,
  Phys.\ Rev.\  D {\bf 65} (2002) 096002;\\
  J.~Bl\"umlein, B.~Geyer and D.~Robaschik,
  Nucl.\ Phys.\  B {\bf 755} (2006) 112;
  arXiv:0706.2478 [hep-ph].
%
\bibitem{THR_ANOM}
  E.~V.~Shuryak and A.~I.~Vainshtein,
  Nucl.\ Phys.\  B {\bf 201} (1982) 141;\\
  A.~P.~Bukhvostov, E.~A.~Kuraev and L.~N.~Lipatov,
  JETP Lett.\  {\bf 37} (1983) 482
  [Pisma Zh.\ Eksp.\ Teor.\ Fiz.\  {\bf 37} (1983\ SPHJA,60,22-32.1984\ ZETFA,87,37-55.1984) 406];\\
  A.~P.~Bukhvostov, G.~V.~Frolov, L.~N.~Lipatov and E.~A.~Kuraev,
  Nucl.\ Phys.\  B {\bf 258} (1985) 601;\\
  P.~G.~Ratcliffe,
  Nucl.\ Phys.\  B {\bf 264} (1986) 493;\\
  I.~I.~Balitsky and V.~M.~Braun,
  Nucl.\ Phys.\  B {\bf 311} (1989) 541;\\
  X.~D.~Ji and C.~h.~Chou,
  Phys.\ Rev.\  D {\bf 42} (1990) 3637;\\
  A.~Ali, V.~M.~Braun and G.~Hiller,
  Phys.\ Lett.\  B {\bf 266} (1991) 117;\\
  J.~Kodaira, Y.~Yasui and T.~Uematsu,
  Phys.\ Lett.\  B {\bf 344} (1995) 348;\\
  J.~Kodaira, Y.~Yasui, K.~Tanaka and T.~Uematsu,
  Phys.\ Lett.\  B {\bf 387} (1996) 855;\\
  J.~Kodaira, T.~Nasuno, H.~Tochimura, K.~Tanaka and Y.~Yasui,
  Prog.\ Theor.\ Phys.\  {\bf 99} (1998) 315;\\
  B.~Geyer, D.~M\"uller and D.~Robaschik,
  Nucl.\ Phys.\ Proc.\ Suppl.\  {\bf 51C} (1996) 106;\\
  D.~M\"uller,
  Phys.\ Lett.\  B {\bf 407} (1997) 314;\\
  V.~M.~Braun, G.~P.~Korchemsky and A.~N.~Manashov,
  Phys.\ Lett.\  B {\bf 476} (2000) 455;
  Nucl.\ Phys.\  B {\bf 597} (2001) 370;
B {\bf 603} (2001) 69.
%
\bibitem{THR_WIL1}
  X.~D.~Ji, W.~Lu, J.~Osborne and X.~T.~Song,
  Phys.\ Rev.\  D {\bf 62} (2000) 094016;\\
  A.~V.~Belitsky, X.~D.~Ji, W.~Lu and J.~Osborne,
  Phys.\ Rev.\  D {\bf 63} (2001) 094012.
%
\bibitem{Zheng:2004ce}
  X.~Zheng {\it et al.}  [Jefferson Lab Hall A Collaboration],
  Phys.\ Rev.\  C {\bf 70} (2004) 065207.
%
\bibitem{LATT1}
  M.~G\"ockeler {\it et al.}, [QCDSF collaboration],
  Phys.\ Rev.\  D {\bf 72} (2005) 054507;\\
  H.~W.~Lin,
  arXiv:0707.3844 [hep-lat].
%
\bibitem{NEG}
  D.~Dolgov {\it et al.}  [LHPC collaboration],
  Phys.\ Rev.\  D {\bf 66} (2002) 034506.
%
\bibitem{KL}
  R.~Kirschner and L.~N.~Lipatov,
  Nucl.\ Phys.\  B {\bf 213} (1983) 122.
%
\bibitem{BV1}
  J.~Bl\"umlein and A.~Vogt,
  Phys.\ Lett.\  B {\bf 370} (1996) 149;\\
  J.~Bl\"umlein, S.~Riemersma and A.~Vogt,
  Nucl.\ Phys.\ Proc.\ Suppl.\  {\bf 51C} (1996) 30.
%
\bibitem{BER}
  J.~Bartels, B.~I.~Ermolaev and M.~G.~Ryskin,
  Z.\ Phys.\  C {\bf 72} (1996) 627.
%
\bibitem{BV2}
  J.~Bl\"umlein and A.~Vogt,
  Phys.\ Lett.\  B {\bf 386} (1996) 350.
%
\bibitem{BV3}
  J.~Bl\"umlein, V.~Ravindran, W.~L.~van Neerven and A.~Vogt,
  arXiv:hep-ph/9806368;\\
  J.~Bl\"umlein and A.~Vogt,
  Phys.\ Rev.\  D {\bf 58} (1998) 014020;
D {\bf 57} (1998) 1.
%
\bibitem{BN1}
  J.~Bl\"umlein and W.~L.~van Neerven,
  Phys.\ Lett.\  B {\bf 450} (1999) 412.
%
\bibitem{UNPOL}
  A.~D.~Martin, R.~G.~Roberts, W.~J.~Stirling and R.~S.~Thorne,
  Eur.\ Phys.\ J.\  C {\bf 35} (2004) 325;\\
  J.~Pumplin, A.~Belyaev, J.~Huston, D.~Stump and W.~K.~Tung,
  JHEP {\bf 0602} (2006) 032;\\
  S.~Alekhin, K.~Melnikov and F.~Petriello,
  Phys.\ Rev.\  D {\bf 74} (2006) 054033;\\
  J.~Bl\"umlein, H.~B\"ottcher and A.~Guffanti,
  Nucl.\ Phys.\  B {\bf 774} (2007) 182;\\
  J.~Bl\"umlein,
  arXiv:0706.2430 [hep-ph].
%
\bibitem{EIC}
C. Aidala et al. [EIC Working Group], A White Paper Prepared for the NSAC 
LPR 2007, {\sf A High Luminosity, High Enery Electron-Ion-Collider}.
%
\bibitem{BB2}
  J.~Bl\"umlein and H.~B\"ottcher,
  Nucl.\ Phys.\  B {\bf 636} (2002) 225.
%
\bibitem{AAC}
  M.~Hirai, S.~Kumano and N.~Saito,
  Phys.\ Rev.\  D {\bf 74} (2006) 014015.
%
\bibitem{GRSV}
  M.~Gl\"uck, E.~Reya, M.~Stratmann and W.~Vogelsang,
  Phys.\ Rev.\  D {\bf 63} (2001) 094005.
%
\bibitem{LSS}
  E.~Leader, A.~V.~Sidorov and D.~B.~Stamenov,
  Phys.\ Rev.\  D {\bf 73} (2006) 034023.
%
\bibitem{FNS}
  D.~de Florian, G.~A.~Navarro and R.~Sassot,
  Phys.\ Rev.\  D {\bf 71} (2005) 094018.
%
\bibitem{AK}
  S.~Atashbar Tehrani and A.~N.~Khorramian,
  JHEP {\bf 0707} (2007) 048.
%
\bibitem{NEUROL}
A. Guffanti, talk at SPIN06, Kyoto, 2006;\\
  J.~Rojo {\it et al.}  [NNPDF Collaboration],
  arXiv:0706.2130 [hep-ph];\\
  L.~Del Debbio, S.~Forte, J.~I.~Latorre, A.~Piccione and J.~Rojo  [NNPDF
                  Collaboration],
  JHEP {\bf 0703} (2007) 039.
%
\bibitem{COMP1}
  V.~Y.~Alexakhin {\it et al.}  [COMPASS Collaboration],
  Phys.\ Lett.\  B {\bf 647} (2007) 8.
%
\bibitem{BB6}
J. Bl\"umlein and H. B\"ottcher, in preparation.
%
\bibitem{KUREK1}
K. Kurek, these proceedings.
%
\bibitem{DELTA_G}
  K.~Kurek,
  arXiv:hep-ex/0607061;\\
  G.~K.~Mallot,
  arXiv:hep-ex/0612055;\\
  S.~Koblitz  [COMPASS Collaboration],
  arXiv:0707.0175 [hep-ex].
%
\bibitem{LATT2}
  Ph.~H\"agler {\it et al.}  [LHPC Collaborations],
  arXiv:0705.4295 [hep-lat];\\
  W.~Schroers,
  Eur.\ Phys.\ J.\  A {\bf 31} (2007) 784;\\
  R.~G.~Edwards {\it et al.}  [LHPC Collaboration],
  Phys.\ Rev.\ Lett.\  {\bf 96} (2006) 052001;\\
  A.~A.~Khan {\it et al.},
  Phys.\ Rev.\  D {\bf 74} (2006) 094508;\\
K. Jansen et al., in preparation.
%
\bibitem{DVCS}
  A.~V.~Belitsky and A.~V.~Radyushkin,
  Phys.\ Rept.\  {\bf 418} (2005) 1 and references therein.
%
\bibitem{LQ}
  F.~Ellinghaus, W.~D.~Nowak, A.~V.~Vinnikov and Z.~Ye,
  Eur.\ Phys.\ J.\  C {\bf 46} (2006) 729.
%
\bibitem{JI}
  X.~D.~Ji,
  Phys.\ Rev.\ Lett.\  {\bf 78} (1997) 610.
%
\bibitem{BB1}
  J.~Bl\"umlein and H.~B\"ottcher,
  Nucl.\ Phys.\  A {\bf 721} (2003) 333.
%
\bibitem{NUFACT}
  M.~L.~Mangano {\it et al.},
  arXiv:hep-ph/0105155.
\end{thebibliography}
\end{document}